\documentstyle[preprint,prb,eqsecnum,aps]{revtex}
\newcommand{\E}{\mathcal{E}}
\renewcommand{\Re}{\mbox{Re }}
\renewcommand{\Im}{\mbox{Im }}
\begin{document}
\draft
\title{Determining parameters of the Neugebauer family of vacuum
spacetimes in terms of data specified on the symmetry axis}
\author{Frederick J.~Ernst}
\address{FJE Enterprises, Rt.~1, Box~246A, Potsdam, NY 13676}
\maketitle
\begin{abstract}
We express the complex potential $\E$ and the metrical fields $\omega$
and $\gamma$ of {\em all} stationary axisymmetric vacuum spacetimes that
result from the application of two successive quadruple-Neugebauer (or
two double-Harrison) transformations to Minkowski space in terms of data
specified on the symmetry axis, which are in turn easily expressed in
terms of multipole moments.  Moreover, we suggest how, in future papers,
we shall apply our approach to do the same thing for those vacuum solutions
that arise from the application of more than two successive transformations,
and for those electrovac solutions that have axis data similar to that of
the vacuum solutions of the Neugebauer family.
\end{abstract}
\section{Introduction}
In recent years considerable interest has been displayed in stationary
axisymmetric solutions of the Einstein-Maxwell equations that are
characterized by axis data of the form
\begin{equation}
\E = \frac{U-W}{U+W} \; , \; \Phi = \frac{V}{U+W} \; ,
\end{equation}
with
\begin{mathletters}
\begin{eqnarray}
U & = & \sum_{a=0}^{n} U_{a} z^{n-a} \; , \\
V & = & \sum_{a=1}^{n} V_{a} z^{n-a} \; , \\
W & = & \sum_{a=1}^{n} W_{a} z^{n-a} \; ,
\end{eqnarray}
\end{mathletters}
where $\E$ and $\Phi$ are the complex potentials of Ernst,\cite{Ernst2}
$z$ is the Weyl canonical coordinate, and the coefficients in the
polynomials are complex constants.  In a long series of papers, V.~Manko
and his associates\cite{Manko} have been evaluating the complex potentials
and metric fields for particular assignments of the axis data.  For
each such assignment, they solve anew Sibgatullin's integral equation
formulation\cite{Sib} of a Riemann-Hilbert problem.  The question naturally
arises ``Could not all these solutions be obtained at once, rather
than in the piecemeal manner employed by Manko et al.?''  Our objective,
which will be partially achieved in the present paper, is to express
the complex potentials and metric fields of {\em all} such solutions in
terms of arbitrarily prescribed axis data of the form indicated above.

To address this question in a systematic way, we shall divide the solution
of the problem into three parts:
\begin{enumerate}
\item
The general solution of the $n=2$ vacuum problem ($V=0$).
\item
The general solution of the $n=2$ electrovac problem.
\item
The general solution of the problem for all values of $n$.
\end{enumerate}
This procedure will enable us to illustrate the basic ideas within the
simplest context (1), after which we shall assemble heavier artillery
to cope with problems (2) and (3).  We already know that it will not be
possible to solve problem (3) completely without resorting to some
numerical work, but the situation is not quite as grim as one might
suppose.  As we solve problems (1) and (2), we shall point out how we
intend to extend the procedures that we have used there to the case
$n>2$.

The vacuum solutions, which are the subject of the present paper,
all belong to the Neugebauer family,\cite{Neu} i.e., they can be
generated from Minkowski space through the action of $n$ successive
quadruple-Neugebauer B\"{a}cklund transformations.  Alternatively,
these solutions can be constructed using $n$ double-Harrison
B\"{a}cklund transformations,\cite{Har} or that Kinnersley-Chitre
transformation\cite{RetzH} that corresponds to the latter
B\"{a}cklund transformation.  While the solutions have been known
for a long time, this is the first attempt of which we are aware
to express everything directly in terms of arbitrarily specified axis
data in the manner of Manko et al.

The complex potential $\E$ of the solution that results from applying
a succession of $n$ quadruple-Neugebauer B\"{a}cklund transformations
to Minkowski space is given by
\begin{equation}
\E = \frac{U-W}{U+W} \; , \label{E}
\end{equation}
where $U$ is the $2n \times 2n$ determinant
\begin{equation}
U = \left| \begin{array}{ccc}
U_{11} & \cdots & U_{1n} \\
\vdots & & \vdots \\
U_{n1} & \cdots & U_{nn}
\end{array} \right| \; ,
\end{equation}
in which occur the $2 \times 2$ submatrices
\begin{equation}
U_{jk} := \left( \begin{array}{cc}
(K_{2k-1})^{j-1} X_{2k-1}r_{2k-1} & (K_{2k})^{j-1} X_{2k}r_{2k} \\
(K_{2k-1})^{j-1} & (K_{2k})^{j-1}
\end{array} \right) \; ,
\end{equation}
where
\begin{equation}
r_{a} := \sqrt{(z-K_{a})^{2}+\rho^{2}} \; .
\end{equation}
The $2n \times 2n$ determinant $W$ is constructed from $U$ by replacing
the $(2n-1)$-st row of the latter determinant by $K_{1}^{n} \; \cdots
\; K_{2n}^{n}$.  It is left for the reader to verify that $\xi := U/W$
is a solution of the complex potential field equation\cite{Ernst1}
\begin{equation}
(\xi\xi^{*}-1) \nabla^{2}\xi = 2\xi^{*} \nabla\xi \cdot \nabla\xi \; ,
\end{equation}
if the constants $K_{a} \; (a=1,\ldots,n)$ are either real or occur in
complex conjugate pairs and the constants $X_{a} \; (a=1,\ldots,n)$
satisfy
\begin{equation}
X_{a} X_{b}^{*} = 1 \text{ when } K_{a} = K_{b}^{*} \; . \label{XX}
\end{equation}
The Kinnersley-Chitre transformations that correspond to various
combinations of real $K$'s and complex conjugate pairs of $K$'s can
be effectively unified into a single {\em complexified K--C
transformation} in which parameters that are real in the case of
an ordinary K--C transformation are replaced by complex valued
parameters.  This approach was first explored by Hauser, who showed
that one can solve the Hauser-Ernst homogeneous Hilbert problem
just as easily for members of the group $SL(2,C)$ as for members of
$SU(1,1)$.\cite{RetzH}  Rather than think of every possible partition
of $K$'s into real ones and complex conjugate pairs as comprising a
different family of solutions, it is more natural and a lot more
convenient to consider these as {\em real cross sections} of one
family of complexified spacetimes,\cite{complex} which we shall dub
{\em the Neugebauer family}, honoring the person who pioneered the
systematic study of members of this family and to whom the determinantal
expressions for the $\E$ potential are due.

The field $U$ is homogeneous of degree $n$ and the field $W$ is
homogeneous of degree $n-1$ in the $r$'s.  From this it follows, in
particular, that the spacetime is asymptotically flat, with a possible
NUT parameter.  On the symmetry axis, where $\rho=0$, one has
$r_{k} = K_{k}-z$, so $U$ and $W$ are, respectively, polynomials
of degree $n$ and $n-1$:
\begin{mathletters}
\begin{eqnarray}
U & = & \sum_{a=0}^{n} U_{a}z^{n-a} \; , \label{u} \\
W & = & \sum_{a=1}^{n} W_{a}z^{n-a} \; . \label{v}
\end{eqnarray}
\end{mathletters}
In particular, $U_{0}$ is the determinant that is constructed from $U$
by the substitution
\begin{equation}
r_{a} \rightarrow -1 \; (a=1,2,\ldots,n) \; .
\end{equation}

When $U_{0} \ne 0$ one can, if one wishes, readjust the common
factor in $U$ and $W$ so that $U_{0} = 1$.  We shall refer to the
resulting constants $U_{a},W_{a} \; (a=1,2,\ldots,n)$ as the
{\em axis data}, the specification of which uniquely determines the
stationary axisymmetric vacuum spacetime.  When $U_{0}=1$, the
real and imaginary parts of the constants $U_{a},W_{a},
\; (a=1,2,\ldots,n)$ are closely connected with the multipole
moments\cite{Simon,Hoenselaers} that describe this asymptotically
flat spacetime.  If one translates along the $z$-axis so that
$\Re U_{1} = 0$, then $iU_{1}$ is a rotation parameter, and $W_{1}$
a (complex) mass parameter, and so on.  The imaginary part of the
latter parameter is associated with the so-called NUT parameter of
the spacetime.

In general $U$ and $W$ can be expressed (up to a common constant factor)
in the respective forms
\begin{mathletters}
\begin{eqnarray}
U & = & -\frac{(-1)^{n}}{(n!)^{2}}\sum_{k_{1},\ldots,k_{2n}=1}^{2n}
\epsilon_{k_{1}\ldots k_{2n}}
\Delta(K_{k_{1}},\ldots,K_{k_{n}}) \Delta(K_{k_{n+1}},\ldots,K_{k_{2n}})
\nonumber \\ & & \hspace{2in} \times
X_{k_{1}}\ldots X_{k_{n}} r_{k_{1}}\ldots r_{k_{n}} \; , \\
W & = & -\frac{1}{(n-1)!(n+1)!}\sum_{k_{1},\ldots,k_{2n}=1}^{2n}
\epsilon_{k_{1}\ldots k_{2n}}
\Delta(K_{k_{1}},\ldots,K_{k_{n-1}}) \Delta(K_{k_{n}},\ldots,K_{k_{2n}})
\nonumber \\ & & \hspace{2in} \times
X_{k_{1}}\ldots X_{k_{n-1}} r_{k_{1}}\ldots r_{k_{n-1}} \; ,
\end{eqnarray}
\end{mathletters}
where
\begin{equation}
\Delta(K_{1},\ldots,K_{n}) := \left| \begin{array}{cccc}
1 & 1 & \cdots & 1 \\
K_{1} & K_{2} & \cdots & K_{n} \\
\vdots & \vdots & & \vdots \\
K_{1}^{n-1} & K_{2}^{n-1} & \cdots & K_{n}^{n-1}
\end{array} \right|
\end{equation}
denotes the determinant of a Vandermonde matrix.  On the symmetry axis,
these formulas reduce to
\begin{mathletters}
\begin{eqnarray}
U & = & -\frac{(-1)^{n}}{(n!)^{2}}\sum_{k_{1},\ldots,k_{2n}=1}^{2n}
\epsilon_{k_{1}\ldots k_{2n}}
\Delta(K_{k_{1}},\ldots,K_{k_{n}})
\Delta(K_{k_{n+1}},\ldots,K_{k_{2n}})
\nonumber \\ & & \hspace{1in} \times
X_{k_{1}} \cdots X_{k_{n}}
(K_{k_{1}}-z) \cdots (K_{k_{n}}-z) \; , \\
W & = & -\frac{1}{(n-1)!(n+1)!}\sum_{k_{1},\ldots,k_{2n}=1}^{2n}
\epsilon_{k_{1}\ldots k_{2n}}
\Delta(K_{k_{1}},\ldots,K_{k_{n-1}})
\Delta(K_{k_{n}},\ldots,K_{k_{2n}})
\nonumber \\ & & \hspace{1in} \times
X_{k_{1}} \cdots X_{k_{n-1}}
(K_{k_{1}}-z) \cdots (K_{k_{n-1}}-z) \; ,
\end{eqnarray}
\end{mathletters}
which are consistent with the axis values of $U$ and $W$ being given by
expressions (\ref{u}) and (\ref{v}), respectively.

\section{The $n=2$ Solution}

The case $n=1$ is well known to correspond to the Kerr-NUT spacetime.
Therefore, we shall concentrate upon the next simplest case, $n=2$,
which was first considered by Kramer and Neugebauer,\cite{KraNeu} and
where
\begin{mathletters}
\begin{eqnarray}
U & = & - (K_{2}-K_{1})(K_{4}-K_{3})
	(X_{1}X_{2}r_{1}r_{2}+X_{3}X_{4}r_{3}r_{4})
	\nonumber \\ & & \hspace{0.5in}
	+ (K_{3}-K_{1})(K_{4}-K_{2})
	(X_{1}X_{3}r_{1}r_{3}+X_{2}X_{4}r_{2}r_{4})
	\nonumber \\ & & \hspace{0.5in}
	- (K_{4}-K_{1})(K_{3}-K_{2})
	(X_{1}X_{4}r_{1}r_{4}+X_{2}X_{3}r_{2}r_{3})
	\label{U} \; , \\
W & = & -\Delta(K_{2},K_{3},K_{4})X_{1}r_{1}
	+\Delta(K_{3},K_{4},K_{1})X_{2}r_{2}
	\nonumber \\ & & \hspace{0.5in}
	-\Delta(K_{4},K_{1},K_{2})X_{3}r_{3}
	+\Delta(K_{1},K_{2},K_{3})X_{4}r_{4} \; . \label{W}
\end{eqnarray}
\end{mathletters}
A simple calculation yields the following expressions for those complex
constants that appear in the axis data:
\begin{mathletters}
\begin{eqnarray}
U_{0} & = & \mbox{}- (K_{1}-K_{2})(K_{3}-K_{4})
(X_{1}X_{2}+X_{3}X_{4}) \nonumber \\
& & \mbox{}+ (K_{1}-K_{3})(K_{2}-K_{4})
(X_{1}X_{3}+X_{2}X_{4}) \nonumber \\
& & \mbox{}- (K_{1}-K_{4})(K_{2}-K_{3})
(X_{1}X_{4}+X_{2}X_{3}) \; , \label{U0} \\
U_{1} & = & (K_{1}-K_{2})(K_{3}-K_{4})
[(K_{1}+K_{2})X_{1}X_{2} + (K_{3}+K_{4})X_{3}X_{4}] \nonumber \\
& & \mbox{}- (K_{1}-K_{3})(K_{2}-K_{4})
[(K_{1}+K_{3})X_{1}X_{3} + (K_{2}+K_{4})X_{2}X_{4}] \nonumber \\
& & \mbox{}+ (K_{1}-K_{4})(K_{2}-K_{3})
[(K_{1}+K_{4})X_{1}X_{4}
+ (K_{2}+K_{3})X_{2}X_{3}] \; , \\
U_{2} & = & \mbox{}- (K_{1}-K_{2})(K_{3}-K_{4})
(K_{1}K_{2}X_{1}X_{2} + K_{3}K_{4}X_{3}X_{4}) \nonumber \\
& & \mbox{} + (K_{1}-K_{3})(K_{2}-K_{4})
(K_{1}K_{3}X_{1}X_{3} + K_{2}K_{4}X_{2}X_{4}) \nonumber \\
& & \mbox{}- (K_{1}-K_{4})(K_{2}-K_{3})
(K_{1}K_{4}X_{1}X_{4} + K_{2}K_{3}X_{2}X_{3}) \; , \\
W_{1} & = & \Delta(K_{2},K_{3},K_{4}) X_{1}
- \Delta(K_{1},K_{3},K_{4}) X_{2} \nonumber \\
& & \mbox{}+ \Delta(K_{1},K_{2},K_{4}) X_{3}
- \Delta(K_{1},K_{2},K_{3}) X_{4} \; , \\
W_{2} & = & \mbox{}-\Delta(K_{2},K_{3},K_{4}) K_{1}X_{1}
+ \Delta(K_{1},K_{3},K_{4}) K_{2}X_{2} \nonumber \\
& & \mbox{}- \Delta(K_{1},K_{2},K_{4}) K_{3}X_{3}
+ \Delta(K_{1},K_{2},K_{3}) K_{4}X_{4} \; . \label{W2}
\end{eqnarray}
\end{mathletters}
We were surprised how easy it was to solve these equations for
the four complex constants $X_{a} \; (a=1,2,3,4)$ in terms of the
$K$'s and the axis data.

\subsection{Determination of $X_{a} \; (a=1,2,3,4)$}

We begin by using the above expressions for $W_{1}$ and $W_{2}$ to
express $X_{4}$ in terms of the other $X$'s.  Then, we use the
expressions for $W_{2}$ and $U_{1}$ to express $X_{2}$ (and $X_{4}$) in
terms of $X_{1}$ and $X_{3}$.  Next, we use the expressions for $W_{2}$
and $U_{0}$ to express $X_{3}$ (as well as $X_{2}$ and $X_{4}$) in terms
of $X_{1}$.  Finally, we use the expressions for $W_{2}$ and $U_{2}$ to
solve for $X_{1}$ (as well as $X_{2}$, $X_{3}$ and $X_{4}$).  The first
and last equations are linear, while the second and third are, perhaps
surprisingly, only quadratic.  Choosing the roots of the quadratic
equations judiciously, we obtain the following expressions for the $X$'s:
\begin{mathletters}
\begin{eqnarray}
{\mathcal D} X_{1} & = &
[U_{2}W_{2}+U_{1}(U_{2}W_{1}-U_{1}W_{2})]
+ (K_{2}+K_{3}+K_{4}) (U_{2}W_{1}-U_{1}W_{2})
\nonumber \\ & & \hspace{0.5in} \mbox{}
- (K_{2}K_{3}+K_{2}K_{4}+K_{3}K_{4}) W_{2}
- (K_{2}K_{3}K_{4}) W_{1} \; , \label{X1} \\
{\mathcal D} X_{2} & = &
[U_{2}W_{2}+U_{1}(U_{2}W_{1}-U_{1}W_{2})]
+ (K_{3}+K_{4}+K_{1}) (U_{2}W_{1}-U_{1}W_{2})
\nonumber \\ & & \hspace{0.5in} \mbox{}
- (K_{3}K_{4}+K_{3}K_{1}+K_{4}K_{1}) W_{2}
- (K_{3}K_{4}K_{1}) W_{1} \; , \\
{\mathcal D} X_{3} & = &
[U_{2}W_{2}+U_{1}(U_{2}W_{1}-U_{1}W_{2})]
+ (K_{4}+K_{1}+K_{2}) (U_{2}W_{1}-U_{1}W_{2})
\nonumber \\ & & \hspace{0.5in} \mbox{}
- (K_{4}K_{1}+K_{4}K_{2}+K_{1}K_{2}) W_{2}
- (K_{4}K_{1}K_{2}) W_{1} \; , \\
{\mathcal D} X_{4} & = &
[U_{2}W_{2}+U_{1}(U_{2}W_{1}-U_{1}W_{2})]
+ (K_{1}+K_{2}+K_{3}) (U_{2}W_{1}-U_{1}W_{2})
\nonumber \\ & & \hspace{0.5in} \mbox{}
- (K_{1}K_{2}+K_{1}K_{3}+K_{2}K_{3}) W_{2}
- (K_{1}K_{2}K_{3}) W_{1} \; , \label{X4}
\end{eqnarray}
\end{mathletters}
where
\begin{equation}
{\mathcal D} := W_{1}(U_{2}W_{1}-U_{1}W_{2})+W_{2}^{2} \; .
\end{equation}
In these expressions, the complex parameters $U_{a}$, $V_{a}$,
$W_{a} \; (a=1,2,3,4)$ have been rescaled so that $U_{0}=1$.
For each choice of the parameters $K_{a} \; (a=1,2,3,4)$,
these equations assign values to the four complex parameters
$X_{a} \; (a=1,2,3,4)$.

\subsection{Determination of $K_{a} \; (a=1,2,3,4)$}

The requirement (\ref{XX}) leads to the following four conditions
(where $U_{0}=1$):
\begin{mathletters}
\begin{eqnarray}
K_{1}+K_{2}+K_{3}+K_{4} & = & - 2 \Re{U_{1}} \; , \label{E1} \\
K_{1}K_{2} + K_{1}K_{3} + K_{1}K_{4}
+ K_{2}K_{3} + K_{2}K_{4} + K_{3}K_{4} & = &
|U_{1}|^{2} - |W_{1}|^{2} + 2\Re U_{2} \; , \\
K_{2}K_{3}K_{4} + K_{1}K_{3}K_{4}
+ K_{1}K_{2}K_{4} + K_{1}K_{2}K_{3} & = &
- 2 \Re(U_{2}U_{1}^{*}-W_{2}W_{1}^{*}) \; , \\
K_{1} K_{2} K_{3} K_{4} & = & |U_{2}|^{2}-|W_{2}|^{2} \; . \label{E4}
\end{eqnarray}
\end{mathletters}
Incidentally, these four relations are equivalent to the single relation
\begin{equation}
|U(z,0)|^{2} - |W(z,0)|^{2}
= |U_{0}|^{2} (K_{1}-z)(K_{2}-z)(K_{3}-z)(K_{4}-z) \; ,
\end{equation}
which one should also be able to deduce directly from the determinantal
expressions for $U$ and $W$, and which can be generalized in a natural
way for $n>2$.\cite{referee1}

{}From the above expressions it is clear that each of the $K$'s satisfies
the quartic equation
\begin{eqnarray}
0 & = & K_{a}^{4} + 2 \Re{U_{1}} K_{a}^{3}
+ (|U_{1}|^{2}-|W_{1}|^{2}+2\Re{U_{2}}) K_{a}^{2} \nonumber \\
& & + 2 \Re{(U_{2}U_{1}^{*}-W_{2}W_{1}^{*})} K_{a}
+ (|U_{2}|^{2}-|W_{2}|^{2}) \; .
\end{eqnarray}
Moreover, because the coefficients of this quartic equation are real,
the solutions $K_{a} \; (a=1,2,3,4)$ are real, or occur in complex
conjugate pairs.

Of course, using a translation along the $z$ axis, we can always
achieve $\Re{U_{1}} = 0$ or $K_{1} + K_{2} + K_{3} + K_{4}=0$.
Cardano's method of solving the quartic equation
\begin{equation}
K_{a}^{4} - A K_{a}^{2} - B K_{a} + C = 0
\end{equation}
then yields
\begin{mathletters}
\begin{eqnarray}
K_{1} & = & \frac{1}{2} \left\{ k + \sqrt{2A-k^{2}+2B/k} \right\}
\; , \label{k1} \\
K_{2} & = & \frac{1}{2} \left\{ k - \sqrt{2A-k^{2}+2B/k} \right\} \; , \\
K_{3} & = & \frac{1}{2} \left\{ -k + \sqrt{2A-k^{2}-2B/k} \right\} \; , \\
K_{4} & = & \frac{1}{2} \left\{ -k - \sqrt{2A-k^{2}-2B/k} \right\} \; ,
\label{k4}
\end{eqnarray}
\end{mathletters}
where
\begin{equation}
k^{2} := \sqrt[3]{T_{+}} + \sqrt[3]{T_{-}} + \frac{2}{3} A \; ,
\end{equation}
\begin{equation}
T_{\pm} := -\frac{1}{27} A (A^{2}-36C) + \frac{1}{2} B^{2}
\pm \frac{1}{18} \sqrt{T_{0}} \; ,
\end{equation}
and
\begin{equation}
T_{0} := 81 B^{4} - 12 B^{2} A (A^2-36C) - 48 C (A^2-4C)^{2} \; .
\end{equation}
In the vacuum case we have the identifications
\begin{mathletters}
\begin{eqnarray}
A & = & |W_{1}|^{2}-|U_{1}|^{2}-2\Re{U_{2}} \; , \\
B & = & - 2 \Re{(U_{2}U_{1}^{*}-W_{2}W_{1}^{*})} \; , \\
C & = & |U_{2}|^{2}-|W_{2}|^{2} \; ,
\end{eqnarray}
\end{mathletters}
which can be generalized easily for electrovac fields.

An alternative way of expressing this result that facilitates a
comparison with Manko et al.\ is
\begin{mathletters}
\begin{eqnarray}
K_{1} & = & \frac{1}{2} \left\{ k
+ (\kappa_{+} + \kappa_{-}) \right\} \; , \label{K1} \\
K_{2} & = & \frac{1}{2} \left\{ k
- (\kappa_{+} + \kappa_{-}) \right\} \; , \\
K_{3} & = & \frac{1}{2} \left\{ -k
+ (\kappa_{+} - \kappa_{-}) \right\} \; , \\
K_{4} & = & \frac{1}{2} \left\{ -k
- (\kappa_{+} - \kappa_{-}) \right\} \; , \label{K4}
\end{eqnarray}
\end{mathletters}
where
\begin{equation}
\kappa_{\pm} := \sqrt{(A - k^{2}/2) \pm 2d} \; ,
\end{equation}
and
\begin{equation}
d := \frac{1}{2} \sqrt{(A-k^{2}/2)^{2}-(B/k)^{2}} \; .
\end{equation}

When the axis data happen to satisfy the relation
\begin{equation}
B := -2 \Re{(U_{2}U_{1}^{*}-W_{2}W_{1}^{*})} = 0 \; ,
\end{equation}
one also has $k = 0$.  Therefore, one must carefully evaluate the
limit of our expressions for the $K$'s as $B \rightarrow 0$, noting
especially that $\lim_{B \rightarrow 0} B/k$ is finite, and
$\lim_{B \rightarrow 0} d = \sqrt{C}$.  The reader can show that the
result obtained this way is consistent with the fact that when $B=0$,
the square of each $K_{a}$ satisfies a quadratic equation,
\begin{equation}
0 = (K_{a}^{2})^{2} - A K_{a}^{2} + C \; ,
\end{equation}
which can be solved directly.  Without further loss of generality
we can express the solution in the form
\begin{mathletters}
\begin{eqnarray}
K_{1} = - K_{2} & = & \frac{1}{2}(\kappa_{+}+\kappa_{-}) \; ,
\label{K1-2} \\
K_{3} = - K_{4} & = & \frac{1}{2}(\kappa_{+}-\kappa_{-}) \; ,
\label{K3-4}
\end{eqnarray}
\end{mathletters}
where
\begin{equation}
\kappa_{\pm} := \sqrt{|W_{1}|^{2}-|U_{1}|^{2}+2(\pm d - \Re{U_{2}})}
\end{equation}
and
\begin{equation}
d := \sqrt{|U_{2}|^{2}-|W_{2}|^{2}} \; .
\end{equation}
The reader can easily check that all equations (\ref{E1}) through
(\ref{E4}) are satisfied by this solution.

In Eqs.\ (\ref{k1}) through (\ref{k4}) or Eqs.\ (\ref{K1}) through
(\ref{K4}) we have expressed $K_{a} \; (a=1,2,3,4)$ explicitly in
terms of the axis data.  Either of these sets of expressions can
be substituted into Eqs.\ (\ref{X1}) through (\ref{X4}) to obtain
$X_{a} \; (a=1,2,3,4)$ explicitly in terms of the axis data.  The
complex potential $\E$ is then given by Eq.\ (\ref{E}) with $U$
and $W$ rendered by Eqs.\ (\ref{U}) and (\ref{W}), respectively.

\section{The Spacetime Metric}

In principle the metric, which is usually written in the form
\begin{equation}
ds^{2} = f^{-1} [e^{2\gamma}(dz^{2}+d\rho^{2}) + \rho^{2} d\phi^{2}]
- f (dt - \omega d\phi)^{2} \; ,
\end{equation}
can be constructed once the complex potential $\E$ is known.  The field
$f := -g_{tt} = \Re \E$ is given by
\begin{equation}
f = \frac{|U|^{2}-|W|^{2}}{|U+W|^{2}} \; .
\end{equation}
Thus, the infinite redshift surface corresponds to
\begin{equation}
|U|^{2} = |W|^{2} \quad \{\text{Infinite Redshift Surface}\},
\end{equation}
while it can be shown that there is a curvature singularity whenever
\begin{equation}
U + W = 0 \quad \{\text{Curvature Singularity}\}.
\end{equation}
The fields $\omega$ and $\gamma$ can be determined, up to respective
integration constants, by solving the differential equations
\begin{mathletters}
\begin{eqnarray}
\omega_{\rho} & = & -\rho f^{-2} \chi_{z} \; , \\
\omega_{z} & = & \rho f^{-2} \chi_{\rho} \; , \\
\gamma_{\rho} & = & \rho \left\{ \E_{\rho}\E_{\rho}^{*}
- \E_{z}\E_{z}^{*} \right\} / (2f)^{2} \; , \\
\gamma_{z} & = & \rho \left\{ \E_{\rho}\E_{z}^{*}+\E_{z}\E_{\rho}^{*}
\right\} / (2f)^{2} \; ,
\end{eqnarray}
\end{mathletters}
where $\chi := \Im{\E}$.

While, in principle, this is straightforward, in practice it is
extraordinarily tedious to calculate the field $\omega$ in this
way.  Even for $n=2$ the number of terms one encounters is enormous.
Modern solution-generating techniques usually provide some alternative
method for determining $\omega$.  We are most familiar with our own
homogeneous Hilbert problem (HHP) formulation, in which the
$H$-potential of Kinnersley and Chitre plays a key role.  This is a
$2 \times 2$ matrix potential, the negative of the real part of which
is just the metric block
\begin{equation}
h = \left( \begin{array}{cc}
f^{-1} \rho^{2} - f \omega^{2} & f \omega \\
f \omega & - f
\end{array} \right) \; ,
\end{equation}
while the lower right element of the matrix $H$ is the complex
$\E$-potential.  Now, the HHP yields not just a formula for $\E$, but
rather a formula for $H$; namely,
\begin{equation}
H = H^{(0)} + \dot{X}_{+}(0) \Omega \; ,
\end{equation}
where, for Minkowski space,
\begin{equation}
H^{(0)} = \left( \begin{array}{cc}
-\rho^{2} & 0 \\
-2iz & 1
\end{array} \right)
\end{equation}
and, generally,
\begin{equation}
\Omega := \left( \begin{array}{cc}
0 & i \\ -i & 0
\end{array} \right) \; .
\end{equation}
The matrix $X_{+}(t)$ is one of the $2 \times 2$ matrices involved
directly in the HHP.  The $\E$ potential is given by
\begin{equation}
\E = H_{LR} = \E^{(0)} + i \dot{X}_{+}(0)^{LL} \; ,
\end{equation}
while
\begin{equation}
f \omega = - \Re{H^{LL}} = f^{(0)}\omega^{(0)}
- \Im{\dot{X}_{+}(0)^{LR}} \; ,
\end{equation}
where $LL$ and $LR$ refer to {\em lower left} and {\em lower right}
matrix elements, respectively.  The point is that one can evaluate
$\dot{X}_{+}(0)^{LR}$ almost as easily as $\dot{X}_{+}(0)^{LL}$.
The field $\omega$ is determined much more easily this way than by
integrating the differential equations for $\omega$.

For the general $n=2$ solution of the Neugebauer family, the procedure
we have outlined yields a result of the form
\begin{equation}
f(\omega-\omega_{0}) = - \Im \left( \frac{\mathcal{N}}{U+W} \right) \; ,
\label{fomega}
\end{equation}
where
\begin{eqnarray}
\mathcal{N} & = & (K_{1}-K_{2})(K_{3}-K_{4})
\nonumber \\ & & \hspace{0.25in} \times
(K_{1}+K_{2}-K_{3}-K_{4})
(Q_{1}Q_{2}-Q_{3}Q_{4}) \nonumber \\
& & - (K_{1}-K_{3})(K_{2}-K_{4})
\nonumber \\ & & \hspace{0.25in} \times
(K_{1}+K_{3}-K_{2}-K_{4})
(Q_{1}Q_{3}-Q_{2}Q_{4}) \label{N} \\
& & + (K_{1}-K_{4})(K_{2}-K_{3})
\nonumber \\ & & \hspace{0.25in} \times
(K_{1}+K_{4}-K_{2}-K_{3})
(Q_{1}Q_{4}-Q_{2}Q_{3}) \; , \nonumber \\
Q_{k} & := & i [X_{k} r_{k} + (K_{k} - z)] \; ,
\label{Q}
\end{eqnarray}
and $\omega_{0}$ is a real constant, the value of which is determined
by the HHP in such a way that $\omega=0$ on the axis.

Finally, one finds that, for all $n=2$ solutions, the field $\gamma$
is given by the simple expression
\begin{equation}
e^{2\gamma} = (|U|^{2}-|W|^{2})/(|U_{0}|^{2}r_{1}r_{2}r_{3}r_{4}) \; ,
\label{gamma}
\end{equation}
where the constant has been chosen so that $\gamma=0$ on the symmetry
axis.\cite{referee2}

In particular, if we specialize to the case
\begin{equation}
U_{1} = -ia \; , \; U_{2} = b \; , \; W_{1} = m \; , \; W_{2} = 0 \; ,
\quad (a,b,m \text{ real})
\end{equation}
where $a^{2} < m^{2}$, we find that
\begin{equation}
X_{1} = X_{3} = - \frac{\kappa_{+}+ia}{m} \; , \;
X_{2} = X_{4} = \frac{\kappa_{+}-ia}{m} \; .
\end{equation}
Substituting these values into our general expressions for $U$, $W$,
and $\mathcal{N}$, and using the notation of Manko et al.,
\begin{equation}
R_{+} := r_{2} \; , \;
R_{-} := r_{1} \; , \;
r_{+} := r_{4} \; , \;
r_{-} := r_{3} \; ,
\end{equation}
we obtain
\begin{mathletters}
\begin{eqnarray}
U & = & \frac{\kappa_{-}^{2}}{m^{2}} \left\{ (m^{2}-2a^{2})
(R_{-}r_{-}+R_{+}r_{+}) + 2ia\kappa_{+} (R_{-}r_{-}-R_{+}r_{+}) \right\}
\nonumber \\ & & \hspace{0.5in}
+ \kappa_{+}^{2} (R_{-}r_{+}+R_{+}r_{-})
-4b (R_{+}R_{-}+r_{+}r_{-}) \; , \\
W & = & \frac{\kappa_{+}\kappa_{-}}{m} \left\{ (m^{2}-a^{2})
(r_{+}+r_{-}-R_{+}-R_{-})
+ \kappa_{+}\kappa_{-} (R_{+}+R_{-}+r_{+}+r_{-})
\right. \nonumber \\ & & \hspace{0.5in} \left.
+ ia [(\kappa_{+}+\kappa_{-})(r_{-}-r_{+})
+ (\kappa_{+}-\kappa_{-})(R_{+}-R_{-})] \right\} \; ,
\end{eqnarray}
\end{mathletters}
the vacuum specialization of the most recent solution published
by Manko et al.  Our way of expressing the field $\omega$ is a good
deal simpler than is theirs; namely,
\begin{equation}
f (\omega - \omega_{0}) = - \Im \left( \frac{\mathcal{N}}{U+W} \right) \; ,
\end{equation}
where $\omega_{0} = -2a$, and
\begin{equation}
\mathcal{N} = 2 \kappa_{+}\kappa_{-} \left\{
\kappa_{-} (Q_{1}Q_{3}-Q_{2}Q_{4}) - \kappa_{+} (Q_{1}Q_{4}-Q_{2}Q_{3})
\right\} \; ,
\end{equation}
where
\begin{mathletters}
\begin{eqnarray}
Q_{1} & = & i \left[ -\left(\frac{\kappa_{+}+ia}{m}\right) R_{-}
+ \frac{1}{2}(\kappa_{+}+\kappa_{-}) - z \right] \; , \\
Q_{2} & = & i \left[ \left(\frac{\kappa_{+}-ia}{m}\right) R_{+}
- \frac{1}{2}(\kappa_{+}+\kappa_{-}) - z \right] \; , \\
Q_{3} & = & i \left[ -\left(\frac{\kappa_{+}+ia}{m}\right) r_{-}
+ \frac{1}{2}(\kappa_{+}-\kappa_{-}) - z \right] \; , \\
Q_{4} & = & i \left[ \left(\frac{\kappa_{+}-ia}{m}\right) r_{+}
- \frac{1}{2}(\kappa_{+}-\kappa_{-}) - z \right] \; .
\end{eqnarray}
\end{mathletters}

\section{Future Extensions}

In this paper we have succeeded in expressing all the $n=2$ members
of the Neugebauer family of solutions of the vacuum field equations
in terms of data prescribed on the symmetry axis, which in turn are
easily related to the multipole moments of the source of the
gravitational field.\cite{Simon,Hoenselaers}  One first evaluates the
$K$'s using Eqs.\ (\ref{k1})--(\ref{k4}) and the $X$'s using Eqs.\
(\ref{X1})--(\ref{X4}).  Then one evaluates $U$ and $W$ using Eqs.\
(\ref{U}) and (\ref{W}), respectively, and $f\omega$ using Eqs.\
(\ref{fomega}), (\ref{N}) and (\ref{Q}).  Of course, $\gamma$ is
given by Eq.\ (\ref{gamma}).

In a future paper we shall include electromagnetic fields, where the
complex potentials have the forms
\begin{equation}
\E = \frac{U-W}{U+W} \; , \; \Phi = \frac{V}{U+W} \; ,
\end{equation}
and $U$ is homogeneous of degree $n$ while $V$ and $W$ are homogeneous
of degree $n-1$ in $r_{1}, \ldots, r_{2n}$.  Electrovac fields of the
type in which we shall be interested have been generated by at least
two techniques, one due to Alekseev\cite{Alekseev} and the other due
to Cosgrove.\cite{Cosgrove}  We are more familiar with Cosgrove's
approach, which, in its usual formulation, produces directly only the
charged Kerr metric with $a^{2}+e^{2} > m^{2}$.  The complexified
Cosgrove transformation, in which the group $SU(2,1)$ is replaced by
$SL(3,C)$ produces a family of complex spacetimes,\cite{complex} the
real cross sections of which are the electrovac spacetimes we shall
study.

After considering the electrovac extension, we shall pass on to the
case $n > 2$.  For all values of $n$, the $K$'s will satisfy the
equation
\begin{equation}
|U|^{2} + |V|^{2} - |W|^{2} = |U_{0}|^{2} \prod_{a=1}^{2n} (K_{a}-z)
\end{equation}
on the symmetry axis.\cite{referee1}  Except in special cases, it
will not be possible to express the $K$'s as algebraic expressions
in the axis data $U_{a}, V_{a}, W_{a} \; (a=1,\ldots,n)$, because,
in general, the $K$'s will be solutions of an algebraic equation of
degree $2n$.  On the other hand, it may be possible to express the
complex potentials $\E$ and $\Phi$ explicitly in terms of the axis
data {\em and} the $K$'s, with the latter parameters determined
numerically from the axis data or the multipole moments.

Long ago Neugebauer gave the $2n \times 2n$ determinants for $U$ and
$W$, but we are not aware of any similar formula for $f\omega$.  Likely
general forms\cite{disclaim} for $\mathcal{N}$ and $\gamma$ can be
guessed from the expressions we have displayed for the case $n=2$:
\begin{mathletters}
\begin{eqnarray}
\mathcal{N} & = & \frac{1}{(n!)^{2}}\sum_{k_{1},\ldots,k_{2n}=1}^{2n}
\epsilon_{k_{1}\ldots k_{2n}}
\Delta(K_{k_{1}},\ldots,K_{k_{n}}) \Delta(K_{k_{n+1}},\ldots,K_{k_{2n}})
\nonumber \\ & & \hspace{0.5in} \times
[(K_{k_{1}}+\ldots+K_{k_{n}}) - (K_{k_{n+1}}+\ldots+K_{k_{2n}})]
Q_{k_{1}}\ldots Q_{k_{n}} \; ,
\end{eqnarray}
and
\begin{eqnarray}
\exp(2\gamma) & = & (|U|^{2}-|W|^{2})/(|U_{0}|^{2} \prod_{k=1}^{2n} r_{k}) \; .
\end{eqnarray}
\end{mathletters}
We regard it as extremely unlikely that any amount of study of the
particular instances considered by Manko et al.\ would have allowed
one to guess a plausible generally applicable form for $\omega$.
That is the principal advantage to considering the general $n=2$ case
rather than particular examples, no matter how interesting those
particular examples might be with respect to potential physical
applications.

It should also be mentioned that, following Neugebauer, one can also
generalize our work by using an arbitrary Weyl metric as the
{\em seed metric} instead of Minkowski space.  One can conceive of
potential physical applications for some of these metrics too.

\section*{Acknowledgment}

This work was supported in part by grant PHY-93-07762 from the
National Science Foundation.


\begin{references}
\bibitem{Ernst2}
F.~J.~Ernst, Phys.~Rev.\ {\bf 168}, 1415 (1968).
\bibitem{Manko}
V.~S.~Manko and N.~R.~Sibgatullin, Phys.~Letters A {\bf 168}, 343 (1992),
V.~S.~Manko and N.~R.~Sibgatullin, Class.~Quantum~Grav.\ {\bf 9}, L87
(1992),
V.~S.~Manko and N.~R.~Sibgatullin, J.~Math.~Phys.\ {\bf 34}, 170 (1993),
V.~S.~Manko, Phys.~Letters A {\bf 181}, 349 (1993),
V.~S.~Manko, J.~Mart\'{\i}n, E.~Ruiz, N.~R.~Sibgatullin and
M.~N.~Zaripov, Phys.~Rev.~D {\bf xx}, xxxx (1994),
V.~S.~Manko, J.~Mart\'{\i}n and E.~Ruiz, Phys.~Rev.~D {\bf xx},
xxxx (1994).
\bibitem{Sib}
N.~R.~Sibgatullin, {\em Oscillations and Waves in Strong Gravitational
and Electromagnetic Fields} (Nauka, Moscow, 1984; English translation:
Springer-Verlag, 1991).
\bibitem{Neu}
G.~Neugebauer, J.~Phys.~A: Math.~Gen.\ {\bf 13}, L19 (1980).
\bibitem{Har}
B.~K.~Harrison, Phys.~Rev.~Lett.\ {\bf 41}, 1197 (1978); Phys.~Rev.\
{\bf D21}, 1695 (1980).
\bibitem{RetzH}
I.~Hauser, in {\em Lecture Notes in Physics, vol.~205}, Springer-Verlag,
Berlin (1984), pp.~128---175.
\bibitem{Ernst1}
F.~J.~Ernst, Phys.~Rev.\ {\bf 167}, 1175 (1968).
\bibitem{complex}
We should like to make it clear that we are not concerned with complex
spacetimes that involve complex coordinates.  Only the parameters are
complex valued.
\bibitem{Simon}
W.~Simon, J.~Math.~Phys.\ {\bf 25}, 1035 (1984).
\bibitem{Hoenselaers}
C.~Hoenselaers and Z.~Perj\'{e}s, Class.~Quantum~Grav.\ {\bf 7}, 1819
(1990).
\bibitem{KraNeu}
D.~Kramer and G.~Neugebauer, Phys.~Lett.\ {\bf A75}, 259 (1980).
\bibitem{referee1}
A referee has pointed out to this author that it is possible
to derive this equation as well as its electrovac generalization
for all $n$ from Sibgatullin's equation
$$
e(\xi)+\tilde{e}(\xi)+2f(\xi)\tilde{f}(\xi) = 0 \; .
$$
\bibitem{referee2}
The author is again grateful to the same referee for pointing out that
a proof of the validity of this formula for {\em all} $n=2$ solutions
can be found in as yet unpublished work of Manko et al.
\bibitem{Alekseev}
G.~A.~Alekseev, Sov.~Phys.~JETP {\bf 32}, 301 (1980).
\bibitem{Cosgrove}
C.~Cosgrove, J.~Math.~Phys.\ {\bf 22}, 2624 (1981).
\bibitem{disclaim}
We emphasize that these forms for $n > 2$ are merely guesses that
will be scrutinized in a later paper.
\end{references}
\end{document}